\documentclass[conference]{IEEEtran}
\usepackage{graphicx}
\usepackage{subfigure}
\usepackage{amsmath,amsthm}
\usepackage{booktabs}
\usepackage{amsfonts}
\usepackage{cases}
\ifCLASSINFOpdf
\else
\fi
\begin{document}
%
\title{Delay-optimal Data Transmission in Renewable Energy Aided Cognitive Radio Networks}


\author{\IEEEauthorblockN{Tian Zhang\IEEEauthorrefmark{2}\IEEEauthorrefmark{1} and
Wei Chen\IEEEauthorrefmark{2}
}
\IEEEauthorblockA{
\IEEEauthorblockA{\IEEEauthorrefmark{2}State Key Laboratory on Microwave
and Digital Communications,
\\
Tsinghua National Laboratory for Information Science and Technology
(TNList) \\
Department of Electronic Engineering, Tsinghua University, Beijing 100084, China
}
\IEEEauthorblockA{\IEEEauthorrefmark{1}School of Information Science and Engineering,
Shandong Normal University, Jinan 250100, China }
\\ Email: tianzhang.ee@gmail.com, wchen@tsinghua.edu.cn}
}


%


\maketitle

\begin{abstract}
Renewable energy powered cognitive radio (CR) network has gained much attention due to its combination of the CR's spectrum efficiency and the renewable energy's \lq\lq green\rq\rq\quad nature. In the paper, we investigate the delay-optimal data transmission in the renewable energy aided CR networks. Specifically, a primary user (PU) and a secondary user (SU) share the same frequency in an area. The SU's interference to the PU is controlled by interference-signal-ratio (ISR) constraint, which means that the ISR at the PU receiver (Rx) should be less than a threshold. Under this constraint, the renewable energy powered SU aims to minimize the average data buffer delay by scheduling the renewable allocations in each slot. A constrained stochastic optimization problem is formulated when the randomness of the renewable arrival, the uncertainty of the SU's data generation, and the variability of the fading channel are taken into account. By analyzing the formulated problem, we propose two practical algorithms that is optimal for two special scenarios. And the two algorithms respectively give an upper and a lower bound for the general scenario. In addition, the availability of the PU's private information at the SU is discussed. Finally, numerical simulations verify the effectiveness of the proposed algorithm.
\end{abstract}


%
\IEEEpeerreviewmaketitle

\section{Introduction}
Cognitive radio (CR) has been an important research area for its superiority in spectrum efficiency. Generally, there are two categories in CR transmissions: underlay and overlay. In the underlay mode, the primary users (PUs) and the secondary users (SUs) can transmit in the same spectrum simultaneously with the guarantee that the SUs' transmissions should not affect the PUs' to a certain extent (e.g., the interference-signal-ratio at the PUs should not exceed some constant). In the overlay mode, the SUs sense the \lq\lq spectrum holes\rq\rq~that the PUs do not utilize periodically, and then transmit through these holes.

Recently, green (i.e., energy-saving and CO2 emission reduction) transmission becomes a vital consideration in communications system design due to severe energy shortage and environmental problems. Under this circumstances, renewable energy, e.g., solar, has been introduced into wireless communications. Renewable energy or energy harvesting\footnote{Energy harvesting technique is capable of converting the energy from the environment (e.g., solar, ambient radio-frequency (RF) signals) into electrical energy. Here energy harvesting is restricted to generate renewable energy from the renewable sources including solar, wind, etc. The RF energy harvesting is excluded due to the considered renewable energy.}aided wireless communications becomes a hot research topic\cite{IEEETVT15:T. Zhang}-\cite{IEEETVT14:P. He L. Zhao S. Zhou and Z. Niu}.

Especially, the renewable energy aided CR network gains much interest because of its assemblage of spectrum efficiency and green energy advantages (e.g., eco-efficiency and potential energy-efficiency)\cite{IEEECOMST:Xueqing Huang Tao Han and Nirwan Ansari}. In \cite{IEEETWC14:S.Park and D.Hong}, an upper bound on the theoretical achievable throughput of energy harvesting SU has been obtained under overlay mode. In \cite{IEEEJSAC15:S. Yin Z. Qu and S. Li}, the \lq\lq harvesting-sensing-throughput\rq\rq~tradeoff is investigated for energy harvesting CR. An optimal single-slot spectrum sensing strategy for throughput optimization has been proposed. In \cite{IEEESensorJournal14:M. Usman and I. Koo}, a hybrid underlay-overlay cognitive radio with energy harvesting is considered, and an access strategy for maximizing the long-term throughput of the system is derived by applying the partially observable Markov decision process framework. In \cite{IEEEWCNC15:K. Kulkarni and A. Banerjee}, a renewable energy aided cooperative CR system, where SU receives unsuccessful packets transmitted by PU and relays them to PU Rx, is considered. The stable throughput region is analyzed. In \cite{IEEEINFOCOM15:S. Gong L. Duan P. Wang}, the robust optimization is performed for energy harvesting CR with the channel and energy harvesting uncertainties.

In the paper, we investigate a renewable energy powered CR, where a PU shares spectrum with an SU. The PU transmits with constant power. The SU transmitter (Tx) is equipped with an energy harvester (e.g., a photovoltaic), and the harvested renewable energy is stored in a battery before usage. The generated data from application layer of the SU Tx is stored in a first-in-first-out (FIFO) data buffer. In each transmission slot, the SU Tx allocates the stored renewable energy for transmitting some data to the SU Rx. To guarantee the PU's transmission QoS, the ISR at the PU Rx should be less than a threshold (i.e., the SU Tx's transmitting power is constrained in each slot). As delay is an important QoS merit (delay-sensitive traffic such as the video increases sharply in wide-spectrum wireless networks, e.g., LTE networks), we focus on the average buffer delay minimization by scheduling the allocated renewable energy in each slot. Accordingly, a constrained stochastic optimization problem is formulated. Next, we analyze the constraints:
In each slot, the allocated power, the transmitted data, and the PU's ISR should not extend the corresponding battery power, the corresponding buffer data, and the ISR threshold, respectively. Based on activity of the constraints, we propose two algorithms. When the renewable energy constraint is inactive (the data constraint or the ISR constraint is active) in all slots, we propose the optimal greedy algorithm. When the renewable energy constraint is active in all slots, we propose an optimal power re-allocation algorithm. Generally, the two algorithms give the upper bound and lower bound, respectively. Additionally, we explore the case that the PU's private information (e.g., the PU's channel gain) is NOT accessible by the SU.
\section{System model and problem formulation}
\begin{figure}[]
\centering
\includegraphics[width=3.5in]{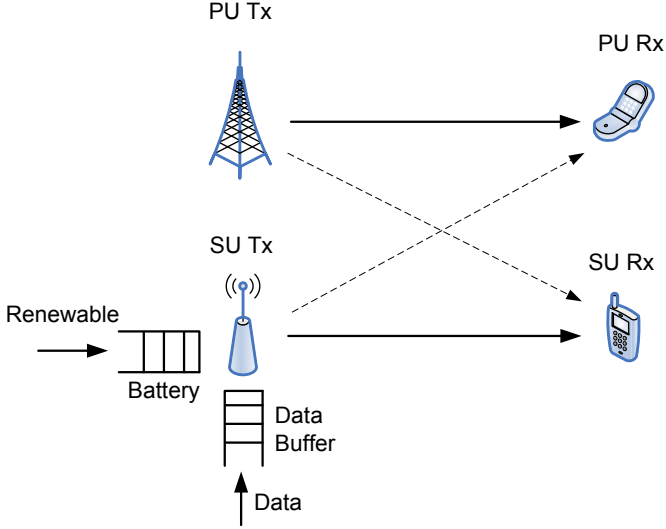}
\caption{Data transmission in renewable energy aided CR networks}
\label{fig_system model}
\end{figure}
Consider a CR network, where a PU co-exits with an SU. Slotted-time model is utilized in the paper, and each slot is with length $\tau$.
 The channel power gains for PU Tx $\&$ PU Rx, PU Tx $\&$  SU Rx, SU Tx $\&$  SU Rx, and SU Tx $\&$ PU Rx during the $n$-th slot are denoted as $g_{11}[n]$, $g_{12}[n]$, $g_{22}[n]$, and $g_{21}[n]$, respectively.
The PU transmits with a fixed power $P_0$. The SU Tx is connected with a renewable energy harvester. The harvested renewable energy is stored in a battery before usage. The data are generated randomly in the application layer of the SU Tx, and wait in an FIFO data buffer. In each slot, some data in the buffer are transmitted to the SU Rx.
Denote the transmitting power of SU in the $n$-th slot as $P[n]$. To guarantee the PU's transmission, the interference-signal-ratio (ISR) at the PU Rx should be less than a given constant, $\rho$, i.e.,
\begin{eqnarray}\label{ISR constraint}
\frac{P[n]g_{21}[n]}{P_0 g_{11}[n]}\le \rho, \forall n.
\end{eqnarray}
Let $E_a[n]$ be the arrived renewable energy at the end of the $n$-th slot (harvested renewable energy during the $n$-th slot). Denote $E[n]$ as the stored energy in the battery at the beginning of the $n$-th slot. Assume that the battery capacity is large enough, we have
\begin{eqnarray}
E[n+1]=E[n]-P[n]\tau+E_a[n]
\end{eqnarray}
Assume that the additive white Gaussian noise at the PU Rx is with zero mean and variance $N_0$. Denote the transmitted data number (in bit) during the $n$-th slot as $R[n]$, then
\begin{eqnarray}
R[n]=\log \Big[\frac{P[n]g_{22}[n]}{P_0g_{12}[n]+N_0}+1\Big]
\end{eqnarray}
Denote the data buffer length at the beginning of $n$-th slot as $Q[n]$, the arrived data at the end of the $n$-th slot (generated data from application layer during the $n$-th slot) as $D_a[n]$. We have
\begin{eqnarray}
Q[n+1]=Q[n]-R[n]+D_a[n]
\end{eqnarray}

The objective is to minimize the average buffer delay over $N$ slots under the ISR constraint by scheduling the renewable allocation in each slot $\left\{P[i]\right\}_{i=1}^{N}$. Accordingly, we have the following problem.

\begin{eqnarray} \label{optimization problem}
\min_{\left\{P[n]\right\}_{n=1}^{N}} \frac{1}{N}\sum\limits_{n =1}^{N+1} Q[n]
\end{eqnarray}

\begin{subequations}
\begin{numcases}{ \mbox{s.t.}}
(\ref{ISR constraint}),\label{ISR constraint InProblem}\\
0\le P[n]\tau \le E[n], \forall n,\label{Renewable constraint}\\
0\le R[n]\le Q[n], \forall n\label{Rate constraint}
\end{numcases}
\end{subequations}
where (\ref{ISR constraint InProblem}) is the ISR constraint, (\ref{Renewable constraint}) denotes the renewable energy constraint, and (\ref{Rate constraint}) is the rate constraint in each slot.
\par
\section{Problem analysis and algorithm design}\label{Section Private information available}
In this section, the constraints in the formulated problem are analyzed. And two optimal algorithms (referred to as the greedy algorithm and the PA algorithm) for special scenarios are proposed thereafter. Moreover, the two algorithms give the upper bound and lower bound respectively. Although the bounds are not the exact solution, they give an important insight into the performance gap.

The three constraints in (\ref{optimization problem}) can be merged as\footnote{
Observe that $\mathcal{C}[n]$ is the function of the optimizing variable $P[n]$.}
\begin{eqnarray}\label{Merged Constraint}
\lefteqn{0\le P[n]}\nonumber\\
&\le &\min\bigg\{\frac{E[n]}{\tau},\frac{\rho P_0 g_{11}[n]}{g_{21}[n]},\big(e^{Q[n]}-1\big)\frac{P_0g_{12}[n]+N_0}{g_{22}[n]}\bigg\}
\nonumber\\
\\
&:=&\mathcal{C}[n]
\end{eqnarray}

\emph{Remark: (\ref{Merged Constraint}) reveals that the renewable energy constraint, the ISR constraint, and the rate constraint can be equivalently transformed to three value-constraints on the allocated power in a slot. Thus, in a slot, only one constraint among the three constraints is active.\footnote{For equal values, consider arbitrary one.}}

Consequently, the SU solves the following optimization problem to get the optimal renewable allocation.
\begin{eqnarray} \label{optimization problem when private available}
\min_{\left\{P[i]\right\}_{i=1}^{N}} \frac{1}{N}\sum\limits_{i =1}^{N+1} Q[i]
\end{eqnarray}
\begin{eqnarray}
\mbox{s.t. }  0\le P[n]\le \mathcal{C}[n], \forall n.
\end{eqnarray}

\subsection{Greedy solution with low-complexity}
Intuitively, we try to transmit as much data as possible, in one slot, so as to minimize the data queue length in data buffer. More data transmission corresponds to more renewable allocation. That is to say, allocate renewable as much as possible in each slot (referred to as \lq\lq greedy renewable allocation\rq\rq) is optimal for minimizing instant data queue length in a slot. Formally, the greedy renewable allocation can be expressed as
\begin{eqnarray}
P[n]=\mathcal{C}[n]
\end{eqnarray}
for $n=1,2,\cdots,N.$
Then there is a natural question\emph{\lq\lq Is the greedy renewable allocation optimal over several slots in average sense?\rq\rq \lq\lq Is the greedy renewable allocation optimal for the formulated problem?\rq\rq}
\subsection{Optimality of the greedy renewable allocation}
With respect to the optimality of the greedy allocation, we have the following lemma.
\theoremstyle{definition} \newtheorem{lemma}{Lemma}
\begin{lemma}
Generally, the greedy renewable allocation is NOT optimal.
\end{lemma}
\begin{IEEEproof}
When the renewable generation process, the data arrival process, and channel state variation are Markov processes, the formulated problem (\ref{optimization problem when private available}) corresponds to $\beta \to 0$ of Lemma 10 in \cite{IEEETVT15:T. Zhang}. Thus, the greedy renewable allocation is not optimal in this case. Thereafter, we arrive at the lemma.
\end{IEEEproof}

\emph{Remark: Although NOT optimal, the greedy renewable allocation is practically useful since its on-line feature and low-complexity. In addition, as the greedy allocation is a feasible solution of (\ref{optimization problem when private available}), it gives an upper bound of the optimal data queue length.}

The greedy renewable allocation is optimal in special circumstances. The following lemma characterizes the scenarios that greedy allocation is optimal.
\begin{lemma}
If
\begin{eqnarray}\label{Sufficient condition for optimality of greedy}
\frac{E[n]}{\tau}\ge \min\Big\{\frac{\rho P_0 g_{11}[n]}{g_{21}[n]},\big(e^{Q[n]}-1\big)\frac{P_0g_{12}[n]+N_0}{g_{22}[n]}\Big\}
\end{eqnarray}
for $n=1,2,\cdots,N$, the greedy renewable allocation is optimal for (\ref{optimization problem when private available}).
\end{lemma}
\begin{IEEEproof}
When (\ref{Sufficient condition for optimality of greedy}) holds, the constraint (\ref{Merged Constraint}) can be reduced as
\begin{eqnarray}\label{1}
P[n]\le \min\Big\{\frac{\rho P_0 g_{11}[n]}{g_{21}[n]},\big(e^{Q[n]}-1\big)\frac{P_0g_{12}[n]+N_0}{g_{22}[n]}\Big\}.
\end{eqnarray}
Define $$R_s[n]:=\log \Big[\frac{\rho P_0 g_{11}[n]}{g_{21}[n]}\frac{g_{22}[n]}{P_0g_{12}[n]+N_0}+1\Big].$$
(\ref{1}) is equivalent to
$R[n]\le \min\Big\{Q[n],R_s[n]\Big\}$.
(\ref{optimization problem when private available}) becomes
\begin{eqnarray}
\min_{\left\{R[i]\right\}_{i=1}^{N}}  \frac{1}{N}\sum\limits_{i =1}^{N+1} Q[i]
\end{eqnarray}
\begin{eqnarray}
\mbox{s.t. }  R[n]\le \min\Big\{Q[n],R_s[n]\Big\}.
\end{eqnarray}
The optimal solution is $$R^*[n]=\min\Big\{Q[n],R_s[n]\Big\};$$ On the other hand, if (\ref{Sufficient condition for optimality of greedy}) is satisfied, the greedy renewable allocation becomes $P[n]=\min\Big\{\frac{\rho P_0 g_{11}[n]}{g_{21}[n]},\big(e^{Q[n]}-1\big)\frac{P_0g_{12}[n]+N_0}{g_{22}[n]}\Big\}$ or $R[n]= \min\Big\{Q[n],R_s[n]\Big\}$ equivalently. Hence we reach the lemma.
\end{IEEEproof}

\emph{Remark:
(\ref{Sufficient condition for optimality of greedy}) gives a sufficient condition for the optimality of the greedy renewable allocation. It is a strong condition since (\ref{Sufficient condition for optimality of greedy}) should be satisfied for arbitrary $n \in [1,\cdots,N]$.}

\emph{Remark:
 $\frac{E[n]}{\tau}\ge \big(e^{Q[n]}-1\big)\frac{P_0g_{12}[n]+N_0}{g_{22}[n]}$ means that the renewable energy is enough for emptying the data buffer in each slot. $\frac{E[n]}{\tau}\ge \frac{\rho P_0 g_{11}[n]}{g_{21}[n]}$ demonstrates that the available renewable energy is more than the ISR upper bound on the SU's transmitting power in each slot. Alternatively, (\ref{Sufficient condition for optimality of greedy}) means that the renewable energy constraint is inactive in all slots.
 In either case, the greedy policy that transmits as much data as possible in each slot is optimal in average sense over slots. }

\subsection{Power re-allocation algorithm}
Since $Q[n]=Q[1]+\sum\limits_{j=1}^{n-1}\Big(D_a[j]-R[j]\Big)$, we have
$$\sum\limits_{n =1}^{N+1} Q[n]=NQ[1]+\sum\limits_{n =1}^{N+1}\sum\limits_{j =1}^{n-1}D_a[j]-\sum\limits_{n =1}^{N+1}\sum\limits_{j =1}^{n-1}R[j].$$ Then (\ref{optimization problem when private available}) is equivalent to the following problem.
\begin{eqnarray} \label{equivalent optimization problem when private available}
\max_{\left\{P[n]\right\}_{n=1}^{N}} \sum\limits_{n =1}^{N} \beta_n\log \Big[\alpha_nP[n]+1\Big]
\end{eqnarray}
\begin{equation}
{ \mbox{s.t.  }} 0\le P[n]\le \mathcal{C}[n],\forall n
\end{equation}
where $\alpha_n=\frac{g_{22}[n]}{P_0g_{12}[n]+N_0}$, $\beta_n=\frac{N+1-n}{N}$.
\par
If
\begin{eqnarray}\label{renewable active}
\frac{E[n]}{\tau}\le \min\bigg\{\frac{\rho P_0 g_{11}[n]}{g_{21}[n]},\big(e^{Q[n]}-1\big)\frac{P_0g_{12}[n]+N_0}{g_{22}[n]}\bigg\}
\end{eqnarray}
for $n=1,2,\cdots,N$, $\mathcal{C}[n]=E[n]$. Then (\ref{equivalent optimization problem when private available}) becomes
\begin{eqnarray} \label{optimization problem with renewable constraint}
\max_{\left\{P[n]\right\}_{n=1}^{N}} \sum\limits_{n =1}^{N} \beta_n\log \Big[\alpha_nP[n]+1\Big]
\end{eqnarray}

\begin{subequations}
\begin{numcases}{ \mbox{s.t.}}
0\le P[n],\forall n\\
\sum\limits_{n=1}^l P[n]\le \sum\limits_{n=0}^{l-1} E_a[n], \forall l
\end{numcases}
\end{subequations}
\par

\emph{Remark: (\ref{renewable active}) denotes that the renewable energy constraint is active in all slots.}

Let $\gamma_n=\frac{1}{\alpha_n\beta_n}$, and assume $\big\{\gamma_{n}\big\}$ being increasing order (the indexes can be arbitrarily
renumbered to comply with this condition). The optimal solution of (\ref{optimization problem with renewable constraint}) can be given by a re-allocation algorithm (i.e., PA algorithm) in Table \ref{Alloc}.
Steps 2.1 - 2.3 are the water-filling procedures that give the optimal power allocation for

\begin{eqnarray} \label{}
\max_{\left\{P[n]\right\}_{n=r}^{k}} \sum\limits_{n =r}^{k} \beta_n\log \Big[\alpha_nP[n]+1\Big]
\end{eqnarray}

\begin{subequations}
\begin{numcases}{ \mbox{s.t.}}
0\le P[n],\forall n\\
\sum\limits_{n=r}^k P[n]=P.
\end{numcases}
\end{subequations}


\begin{table}[]
   \caption{}\label{Alloc}
   \centering
    \begin{tabular}{lcl}
     \toprule
     \textbf{Power re-Allocation (PA) algorithm: $Alloc()$} \\
     \midrule
     Input:  $N$, $\{E_a[n]\}_{n=1}^N$, $\{\alpha_n\}_{n=1}^N$\\
    Step 1:  $Alloc(1)=P^*[1]=E_a[0]$.\\
    Step 2: \\
    ~for $k=2:N$\\
             ~~$\Big\{P^{'}[i]\Big\}_{i=1}^{k-1}=Alloc(k-1)$\\
              ~~~for $r=k:-1:1$\\
              ~~~~$P=\sum_{i=r}^{k-1}P^{'}[i]+E_a[k-1]$\\
              ~~~~Step 2-1: $\Delta=0$, $P_M=P^*=P$, $q=r$;\\
              ~~~~Step 2-2: $\Delta=\Delta+\beta_q$, $P^*=P^*-\Big(\gamma_{q+1}-\gamma_{q}\Big)\Delta$, $q=q+1$;\\
              ~~~~Step 2-3: if $P^* >0$ and $q\le k$, $P_M=P^*$, repeat Step 2-2; \\
            ~~~~~~~~~~~~~~~else $q^*=q-1$, $P[q^*]=\frac{\beta_{q^*}}{\Delta}P_M$.\\
           ~~~~~~~~~~~~~~~end if\\
            ~~~~\begin{subequations}
       \begin{numcases}{P^*[q]=}
         \Big[\frac{P[q^*]}{\beta_{q^*}}+\gamma_{q^*}-\gamma_q\Big]\beta_q,r\le q\le q^*\nonumber\\
         0,q^*< q \le k.\nonumber
       \end{numcases}
     \end{subequations}\\
              ~~~~if $r>1$, $q_e=\max\Big\{q\Big|P^{'}[q]>0,1\le q\le r-1\Big\}$\\
              ~~~~else $q_e=1$.\\
              ~~~~end if\\
              ~~~~if $\frac{1}{\alpha_{q^*}\beta_{q^*}}+\frac{P^{*}[q^*]}{\beta_{q^*}}\ge \frac{1}{\alpha_{q_e}\beta_{q_e}}+\frac{P^{'}[q_e]}{\beta_{q_e}}$\\
              ~~~~$Alloc(k)=\Big\{P^{'}[1],\cdots,P^{'}[r-1],P^{*}[r],\cdots,P^{*}[k]\Big\}$, break.\\
              ~~~~end if\\
          ~~end for\\
              ~end for\\
     Output: $\{P^*[n]\}_{n=1}^N=Alloc(N)$\\
     \bottomrule
    \end{tabular}
   \end{table}
\par
\emph{Remark:
The PA algorithm gives the optimal solution for a relaxed version of (\ref{optimization problem when private available}). The constraint $ 0\le P[n]\le \mathcal{C}[n],\forall n$ is relaxed to be $ 0\le P[n]\le \frac{E[n]}{\tau},\forall n$. Although NOT optimal generally, it incurs a lower bound of the optimal data queue length.}

\emph{Remark: The PA algorithm derives the optimal power allocation of (\ref{optimization problem when private available}) in special scenarios.}
\par
\emph{Remark: The PA algorithm is an off-line algorithm.}

In the following, we give two numerical examples to illustrate the PA algorithm's executing process.

\begin{figure}[tbp]
\centering
\subfigure[Slot 1]{\includegraphics[width=1.0in]{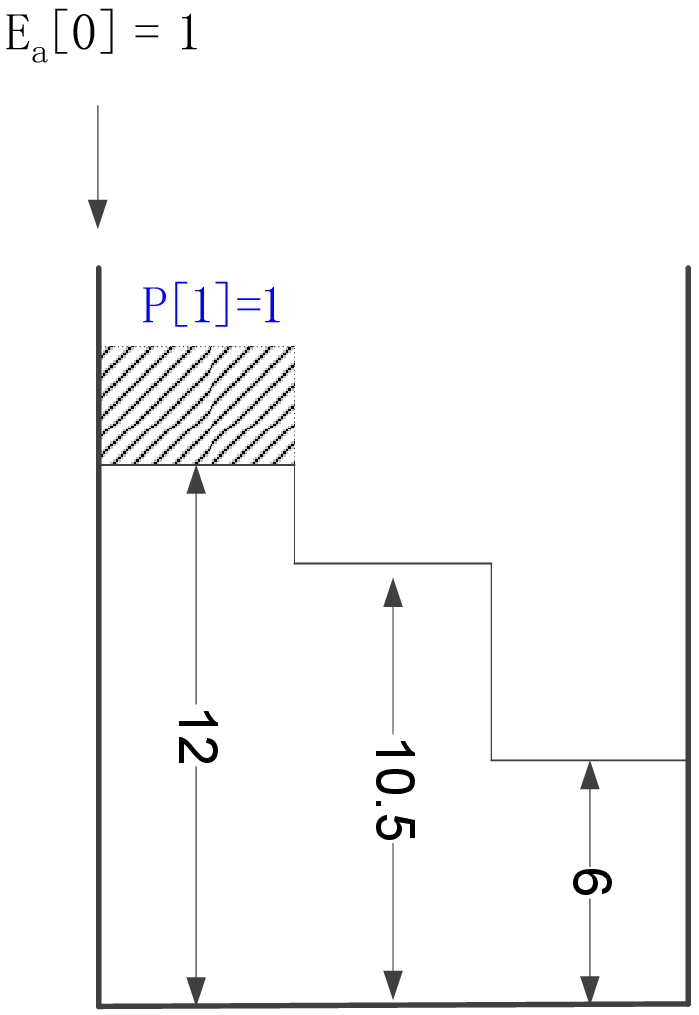}}
\subfigure[Slot 2]{\includegraphics[width=0.9in]{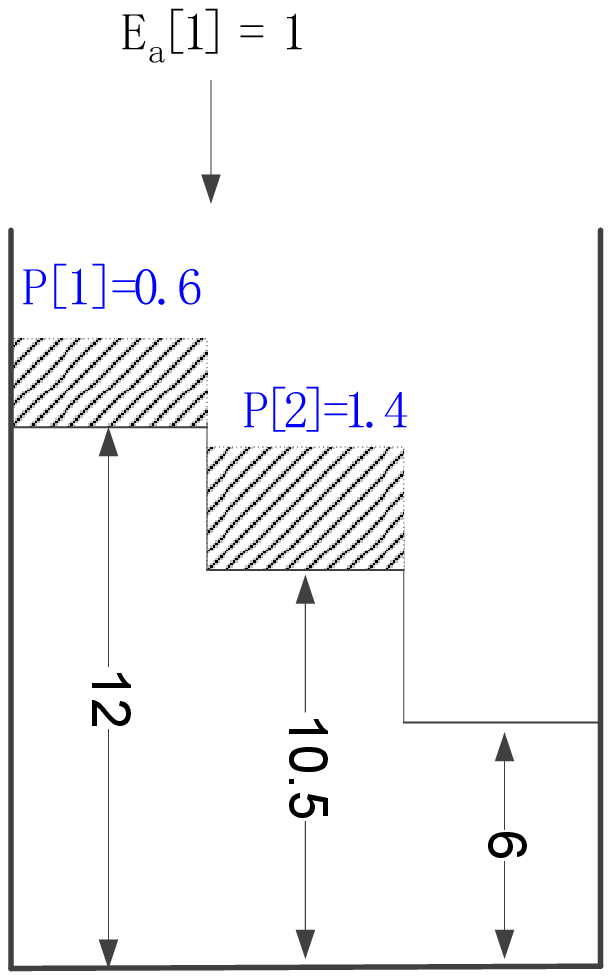}}
\subfigure[Slot 3]{\includegraphics[width=1.0in]{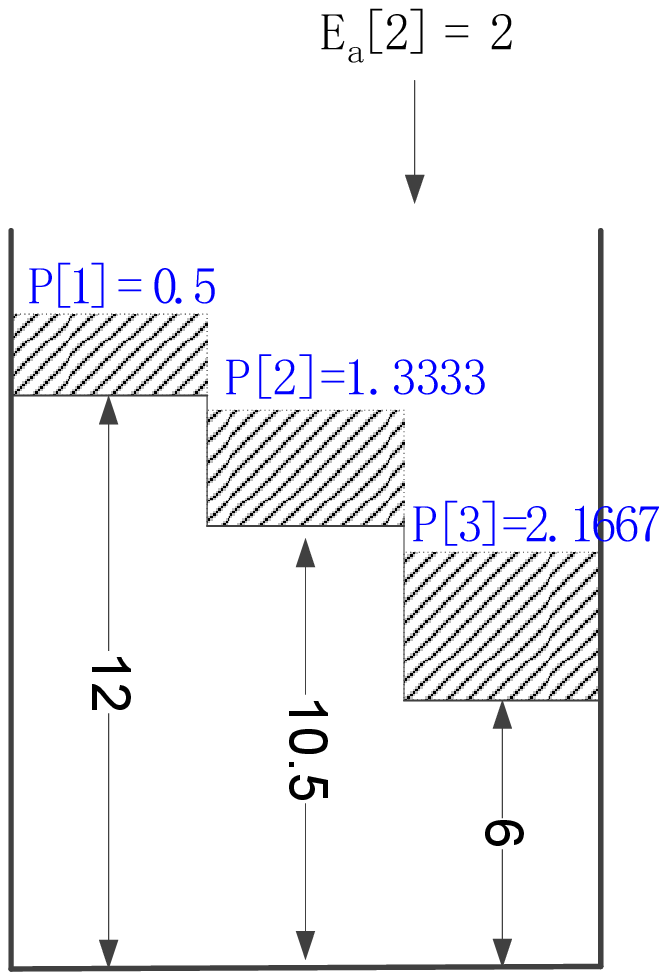}}
\caption{PA algorithm for Numerical Example 1}
\label{Example1}
\end{figure}

\begin{figure}[tbp]
\centering
\subfigure[Slot 1]{\includegraphics[width=1.0in]{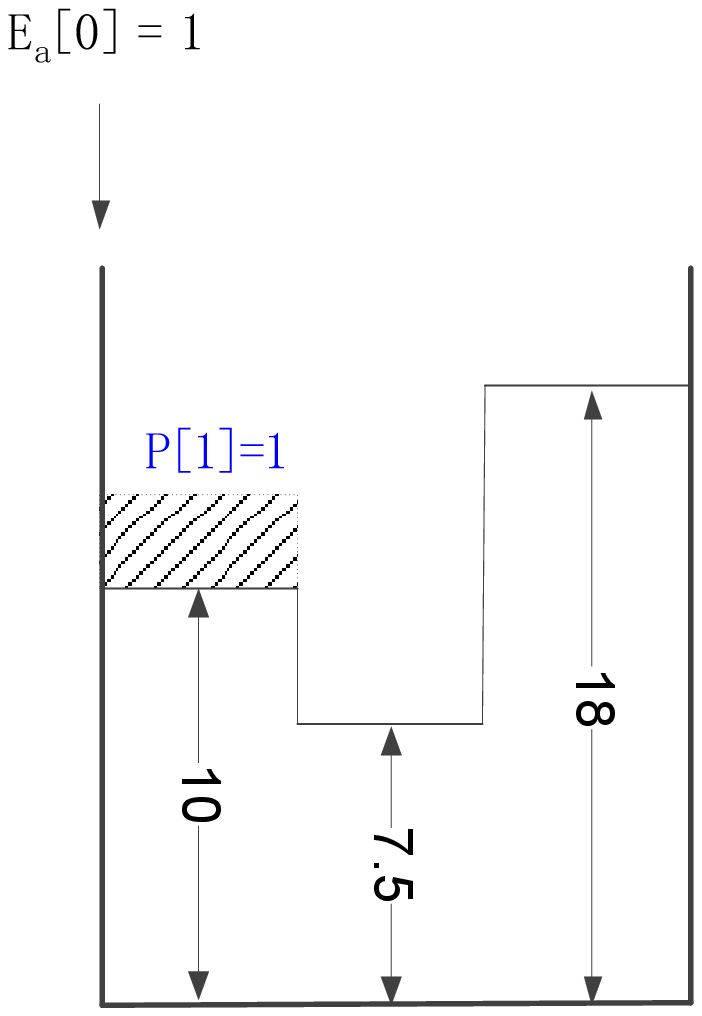}}
\subfigure[Slot 2]{\includegraphics[width=0.9in]{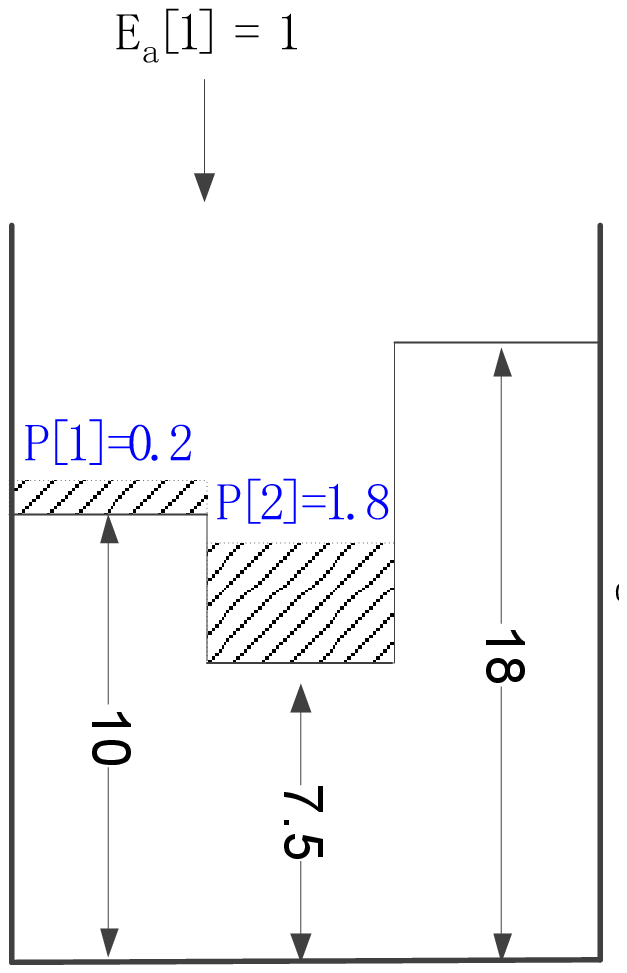}}
\subfigure[Slot 3]{\includegraphics[width=0.9in]{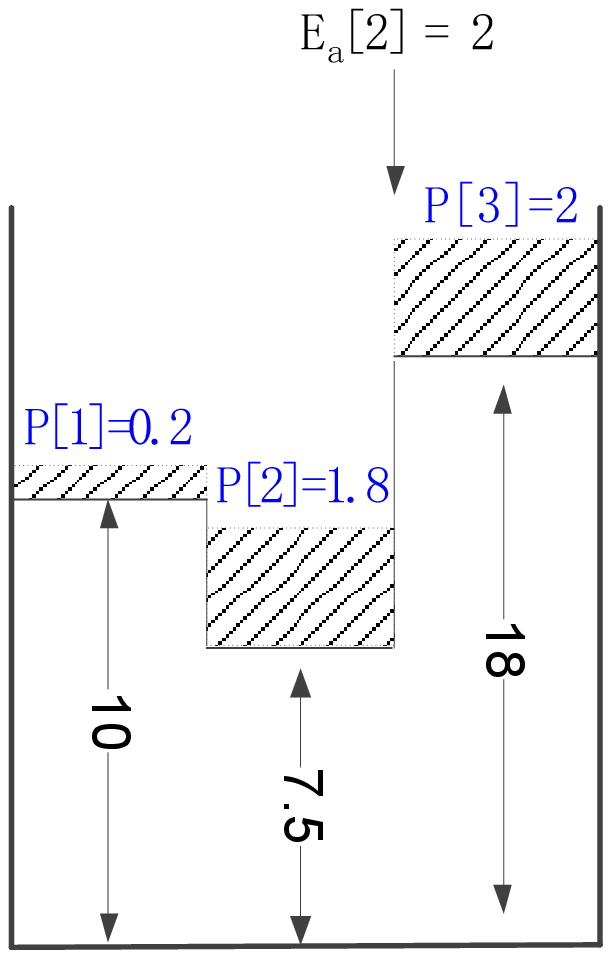}}
\caption{PA algorithm for Numerical Example 2}
\label{Example2}
\end{figure}

\emph{Numerical examples:} Let $N=3$, $Ea=[1~ 2~ 1]$, the initial renewable energy $E_a[0]=E[0]=1$. Fig. \ref{Example1} demonstrates the power allocation of the PA algorithm for $\alpha=[1/12~ 1/7~ 1/2]$ (Numerical Example 1), $Alloc(1)=1$, $Alloc(2)=[0.6000~ 1.4000]$,
$Alloc(3)=[0.5000~ 1.3333~ 2.1667]$. For slot 1, the renewable at slot 1 is the initial renewable energy $E[0]=1$, all renewable energy is allocated, $P[1]=1$. For slot 1 \& slot 2, the maximal renewable energy at slot 2 is $E[0]+Ea[1]=1+1=2$. As the channel condition for slot 2 ($\alpha[2]$) is better enough than slot 1 ($\alpha[1]$), then the power allocation is adjusted and the re-allocation occurs. More power is allocated at slot 2, $P[1]=0.4$, $P[2]=1.6$. For slot 1 \& slot 2 \& slot 3, the maximal available renewable energy is $E[0]+Ea[1]+Ea[2]=1+1+2=4$. Since the channel condition for slot 3 ($\alpha[3]$) is better enough than slot 1 and slot 2, the power is re-allocated once again, more power is utilized for data transmission at slot 3, $P[1]=0.5000$, $P[2]=1.3333$, $P[3]=2.1667]$.
Fig. \ref{Example2} shows the power allocation of the PA algorithm for $\alpha=[1/10~ 1/5~ 1/6]$ (Numerical Example 2), $Alloc(1)=1$, $Alloc(2)=[0.2000~ 1.8000]$,
$Alloc(3)=[0.2000~ 1.8000~ 2.0000]$. The explanations for slot 1 and slot 1 \& slot 2 are same as former example. For slot 1 \& slot 2 \& slot 3, As the channel condition for slot 3 ($\alpha[3]$) is NOT better enough than slot 1 and slot 2, then no re-allocation occurs. Power allocation of slot 1 and slot 2 remains static as that for slot 1 \& slot 2, and the newly arrived power at slot 3 ($E_a[3]=2$) is totally allocated for slot 3.

\section{Discussion on the availability of the PU's private information at the SU }
The SU's renewable allocation constraint is related to $P_0$, $g_{11}$, etc. Hence there are two scenarios:
\begin{itemize}[]
\item The private information of the PU, e.g., $P_0$, $g_{11}$, can be obtained by the SU.
\item The private information of the PU is unavailable at the SU.
\end{itemize}
We have studied the scenario that the PU's private information is available at the SU in Section \ref{Section Private information available}. In this section, we investigate the problem \emph{How can we do when the private information is unavailable?}

When the private information of the PU (e.g., $P_0$, $g_{11}$) is unavailable, the SU can NOT compute the second item $\frac{\rho P_0 g_{11}[n]}{g_{21}[n]}$ in (\ref{Merged Constraint}). In contrast, $P_0g_{12}[n]+N_0$ can be obtained at the SU Rx and can be fed back to the SU Tx. The third one $\big(e^{Q[n]}-1\big)\frac{P_0g_{12}[n]+N_0}{g_{22}[n]}$ can be obtained.
Consider the following relaxed problem

\begin{eqnarray} \label{relaxed optimization problem when private unavailable}
\min_{\left\{P[n]\right\}_{n=1}^{N}} \frac{1}{N}\sum\limits_{n =1}^{N+1} Q[n]
\end{eqnarray}
\begin{eqnarray}
\mbox{s.t. }  P[n]&\le& \min\bigg\{\frac{E[n]}{\tau},\big(e^{Q[n]}-1\big)\frac{P_0g_{12}[n]+N_0}{g_{22}[n]}\bigg\}
\nonumber\\
&:=&\mathcal{F}[n].
\end{eqnarray}
Since $\mathcal{F}[n]\ge \mathcal{C}[n]$ for arbitrary $n$, (\ref{relaxed optimization problem when private unavailable}) is a relaxed version of (\ref{optimization problem}). That is to say, the optimal data queue length of (\ref{relaxed optimization problem when private unavailable}) gives a lower bound for that of (\ref{optimization problem}). That is to say, even without the PU's private information, the SU can derive a lower bound on the average buffer length by solving a relaxed problem. Furthermore, regarding the relaxed problem (\ref{relaxed optimization problem when private unavailable}), we have the following results.
\begin{lemma}
For problem (\ref{relaxed optimization problem when private unavailable}), when
$$\frac{E[n]}{\tau}\ge \big(e^{Q[n]}-1\big)\frac{P_0g_{12}[n]+N_0}{g_{22}[n]}$$ for $n=1,\cdots,N$, the greedy allocation is optimal; If $$\frac{E[n]}{\tau}< \big(e^{Q[n]}-1\big)\frac{P_0g_{12}[n]+N_0}{g_{22}[n]}$$ for $n=1,\cdots,N$, the PA algorithm gives the optimal allocation.
\end{lemma}
\begin{IEEEproof}
The lemma can be verified by derives in Section \ref{Section Private information available}. The detailed proofs are omitted for brevity.
\end{IEEEproof}
\section{Numerical results}
In this section, simulations are carried out to illustrate the average data buffer length performance of the PA algorithm and the greedy algorithm. Unless otherwise specified, $N=3$, $\tau=1$, $\rho=0.1$, $\frac{P_0g_{11}}{g_{21}}=[100~ 420~ 200]$.

\begin{figure}[]
\centering
\includegraphics[width=3.5in]{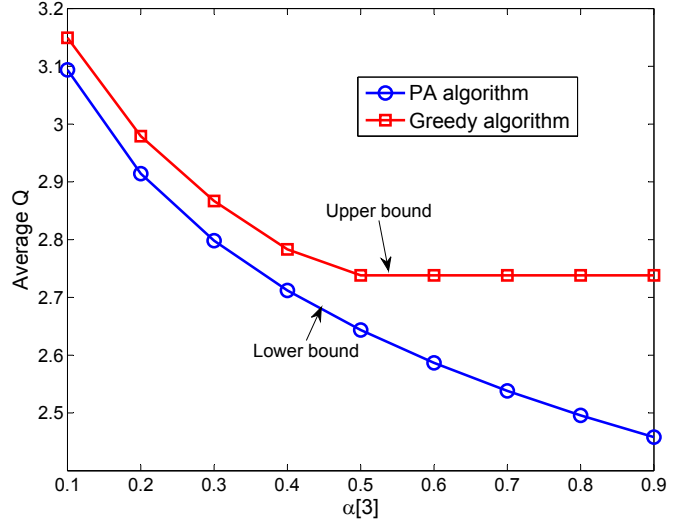}
\caption{Average data buffer length v.s. channel condition}
\label{fig_sim1}
\end{figure}

Fig. \ref{fig_sim1} plots the average buffer length performance v.s. the channel condition $\alpha[3]$ for the greedy algorithm and the PA algorithm, respectively. We have set $\alpha=[1/80~ 1/30~ \alpha[3]]$, the initial renewable energy $E[0]=E_a[0]=12$, $E_a=[20~ 25~ 18]$, the initial data $Q[0]=1$, $D_a=[1~ 1~ 3]$. It can be observed that the average buffer length decreases with the channel condition improvement (increase of $\alpha[3]$) for the PA algorithm. In contrast, for the greedy algorithm, the buffer length decreases sharply in small value region of $\alpha[3]$ and remains almost constant when $\alpha[3]>0.5$. In addition, for bad channel conditions (small value region of $\alpha[3]$) the gap between the PA algorithm and the greedy is apparently smaller than that for the good channel conditions (the lager value region).
Generally, the channel state improvement results in more data transmission given renewable energy arrival. For the PA algorithm, larger value of $\alpha[3]$ incurs the re-allocation of the renewables at slot 3. Specifically, when $\alpha=0.1,~ 0.2$, the PA algorithm produces $ [5.2000~ 26.8000~ 25.0000]$, i.e., the renewable re-allocation happens at slot 2 only. When $\alpha[3]\ge 0.3$, the re-allocation happens at both slot 2 and slot 3. Furthermore, with the increase of $\alpha$, more power is allocated to slot 3 (e.g., the power allocations are $[5.1667~ 26.7778~ 25.0556]$ and $[4.0556~ 26.0370~ 26.9074]$ for $\alpha[3]=0.3$ and $\alpha[3]=0.9$, respectively). Since the channel condition of slot 3 is the best among the there slots, more data can be transmitted. Thus, the average buffer length decreases with the increase of $\alpha[3]$.
For the greedy algorithm,
when $\alpha[3]$ is small, the ISR constraint (\ref{ISR constraint InProblem}) and the renewable constraint (\ref{Renewable constraint}) are active constraints for all the three slots (especially for slot 3). Consequently, the average buffer length decreases with the increase of $\alpha[3]$. Once $\alpha[3]$ is larger enough, the active constraint becomes the rate constraint (\ref{Rate constraint}). That is to say, the transmitted data is determined by the buffer queue length at the beginning of the 3-th slot and it is irrelevant to $\alpha[3]$. Then the average buffer length remains static (In the greedy algorithm, $\alpha[3]$ does NOT affect the data transmissions of slot 1 and slot 2).
When $\alpha[3]>0.5$, the average buffer length for the PA algorithm decreases and remains static for the greedy algorithm. Hence the gap between the two algorithms becomes apparently larger.

\begin{figure}[]
\centering
\includegraphics[width=3.5in]{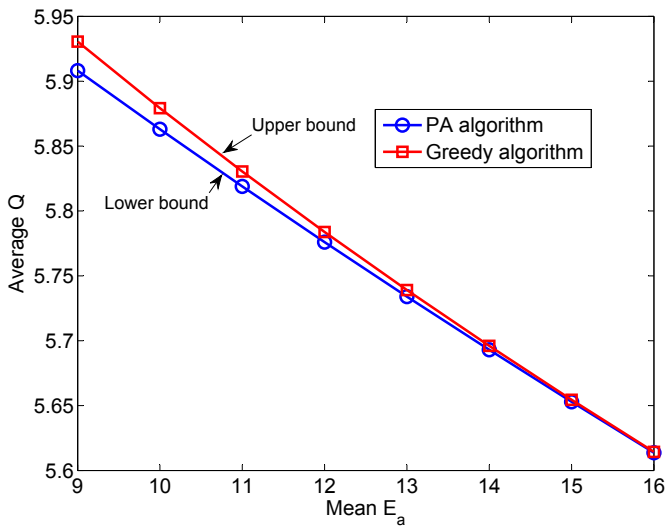}
\caption{Average data buffer length w.r.t. mean energy arrival}
\label{fig_sim2}
\end{figure}

Fig. \ref{fig_sim2} draws the average buffer length performance regarding the mean renewable energy arrival, $\bar{E}_a$, for the PA algorithm and greedy algorithm, respectively. In the simulation, $\alpha=[1/15~ 1/16~ 0.8]$, the initial renewable energy $E[0]=E_a[0]=1/4*\bar{E}_a$, $E_a=[1/4*\bar{E}_a~ 2/4*\bar{E}_a~ 0]$, the initial data $Q[0]=2$, and $D_a=[1~ 2~ 5]$. From the figure, we can find that with the increase of $\bar{E}_a$, the mean buffer length decreases for both algorithms, and the performance gap becomes indistinct (almost zero finally).
For the PA algorithm, since the channel conditions of slot 1 and slot 2 are similar, no power reallocation happens between slot 1 and slot 2. The power re-allocation occurs between slot 2 and slot 3, and the re-allocation becomes weaker with the increase of $\bar{E}_a$. For example, the power allocation is $[2.2500~ 0~ 6.7500]$ for $\bar{E}_a=9$, i.e., the stored renewable energy for slot 2 is totally re-allocated to slot 3, $P[2]=0$ \& $P[3]=E[3]=E_a[1]+E_a[2]$. In contrast, for $\bar{E}_a=16$, the power allocation is $[4.0000~ 3.5000~ 8.5000]$. $P[2]=87.5\%*E_a[1]$ \& $P[3]=12.5\%*E_a[1]+E_a[2]$ and only $0.5/4=12.5\%$ is re-allocated. Regarding the greedy algorithm, renewable energy constraint (i.e., (\ref{Renewable constraint})) is the active constraint. Then, the stored renewable energy is totally utilized in each slot, $P[n]=E[n]=E_a[n-1]$.
When $\bar{E}_a$ increases, for both algorithms, the allocated power in each slot increases, more data is transmitted. Then the average buffer length decreases. Meanwhile, with the increase, the PA allocation approaches the greedy allocation. Thus, the average buffer length gap becomes minor.

\begin{figure}[]
\centering
\includegraphics[width=3.5in]{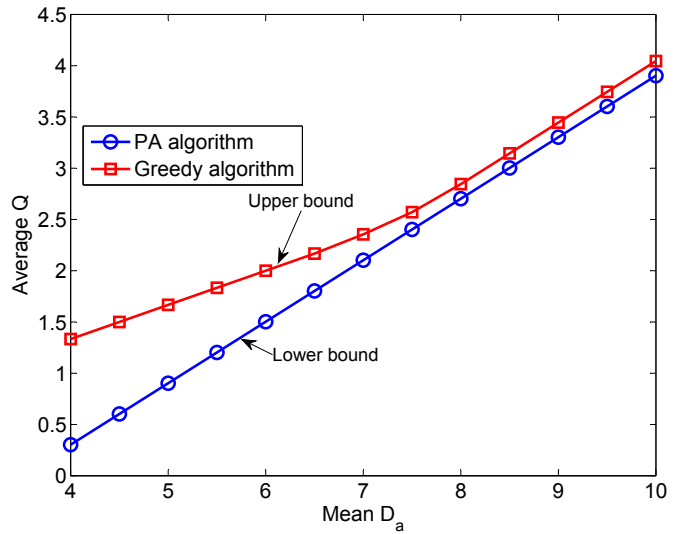}
\caption{Average data buffer length w.r.t. mean data arrival}
\label{fig_sim3}
\end{figure}

Fig. \ref{fig_sim3} illustrates the average buffer length performance of the PA algorithm and the greedy algorithm regarding the mean data arrival, $\bar{D}_a$. In the simulation, $\alpha=[1/15~ 1/3~ 0.5]$, $E[0]=E_a[0]=8$, $E_a=[12~ 10~ 1]$, $Q[0]=0$, and $Da=[0.3*\bar{D}_a~ 0.2*\bar{D}_a~ 0.5*\bar{D}_a]$. It can be seen that the average buffer length increase with the increase of $\bar{D}_a$ according to linear relationship for the PA algorithm. Concerning the greedy algorithm, the increase is slow first and fast then.
The average buffer length gap between the two algorithms becomes smaller in small value region of $\bar{D}_a$ and remains constant in large value region.
It can be explained as follows: The PA algorithm is irrelevant to the mean data arrival, and forasmuch, the produced power allocation remains as $[7.8000~ 12.2000~ 10.0000]$, and the transmitted data $R=R[1]+R[2]+R[3]$ is constant with respected to $\bar{D}_a$. The average buffer length $\bar{Q}=\bar{D}_a-\frac{R}{3}$. Hence the average buffer length varies linearly with $\bar{D}_a$. With regard to the greedy algorithm, since the initial data is zero, the rate constraint is active and the power allocation in slot 1 is zero. When $\bar{D}_a$ is small ($\bar{D}_a\le 6.5$), for slot 2 and slot 3, the rate constraint is active. Hence the increment is slow.
When $\bar{D}_a=7$, $7.5$, renewable energy constraint is active for slot 2 and rate constraint is active for slot 3. The increase is moderate.
When $\bar{D}_a\ge 8$, the renewable constraint is active for slot 2 and slot 3. Then the increase becomes linear. Combing above analysis of the two algorithms, we can explain why the gap becomes smaller first and remains static then.

\section{Conclusion}
Delay-optimal data transmission of CR networks in the presence of renewable energy is studied in the paper. Considering the data generation, renewable energy arrival, and wireless channel state, the SU's renewable energy allocation in each slot is scheduled to minimize the average data buffer length with ISR constraint. By formulating a constrained stochastic optimization problem and corresponding theoretical analysis, we propose two practical algorithms: the on-line greedy algorithm and the off-line PA algorithm. The two algorithms are respectively optimal in certain special conditions. Meanwhile, the greedy algorithm gives an upper bound on the average data buffer length, and the PA algorithm derives a lower bound.
Additionally, the case that SU can NOT get the PU's private information is also discussed. Numerical results demonstrate that the greedy algorithm and the PA algorithm can produce effective upper \& lower bounds on the average buffer length and give important insights into the performance thereafter.

%
%




\begin{thebibliography}{1}
%
%

\bibitem{IEEETVT15:T. Zhang}
T. Zhang, W. Chen, Z. Han, and Z. Cao, \lq\lq A cross-layer perspective on energy harvesting aided green communications over fading channels,\rq\rq \emph{ IEEE Trans. Veh. Technol.}, vol. 64, no. 4, pp. 1519 - 1534, Apr. 2015.
\bibitem{IEEEJSAC15:Roy D. Yates and Hajar Mahdavi-Doost}
R. D. Yates and H. M.-Doost, \lq\lq Energy harvesting receivers: Packet sampling
and decoding policies,\rq\rq \emph{ IEEE J. Sel. Area Comm.}, vol. 33, no. 3, pp. 558 - 570, Mar. 2015.
\bibitem{IEEETVT14:P. He L. Zhao S. Zhou and Z. Niu}
P. He, L. Zhao, S. Zhou, and Z. Niu, \lq\lq Recursive waterfilling for wireless links with
energy harvesting transmitters,\rq\rq \emph{ IEEE Trans. Veh. Technol.}, vol. 63, no. 3, pp. 1232 - 1241, Mar. 2014.
\bibitem{IEEECOMST:Xueqing Huang Tao Han and Nirwan Ansari}
X. Huang, T. Han, and N. Ansari, \lq\lq On green energy powered cognitive radio networks\rq\rq, \emph{IEEE Commun. Surv. \& Tut.}, vol. 17, no. 2, pp. 827 - 842, 2nd-quar. 2015.


\bibitem{IEEETWC14:S.Park and D.Hong}
S. Park and D. Hong, \lq\lq Achievable throughput of energy harvesting cognitive radio networks,\rq\rq \emph{ IEEE Trans. Wireless Commun.}, vol. 13, no. 2, pp. 1010 - 1022, Feb. 2014.

\bibitem{IEEEJSAC15:S. Yin Z. Qu and S. Li}
S. Yin, Z. Qu, and S. Li, \lq\lq Achievable throughput optimization in energy harvesting cognitive radio systems,\rq\rq \emph{ IEEE J. Sel. Area Comm.}, vol. 33, no. 3, pp. 407 - 422, Mar. 2015.

\bibitem{IEEESensorJournal14:M. Usman and I. Koo}
M. Usman and I. Koo, \lq\lq Access strategy for hybrid underlay-overlay cognitive radios with energy harvesting,\rq\rq \emph{ IEEE Sens. J.}, vol. 19, no. 9, pp. 3164 - 3173, Sept. 2014
\bibitem{IEEEWCNC15:K. Kulkarni and A. Banerjee}
K. Kulkarni and A. Banerjee, \lq\lq Stable throughput tradeoffs in cognitive radio networks with cooperating rechargeable nodes,\rq\rq \emph{ Proc. IEEE WCNC'15}, New Orleans, LA, Mar. 2015.

\bibitem{IEEEINFOCOM15:S. Gong L. Duan P. Wang}
S. Gong, L. Duan, and P. Wang, \lq\lq Robust optimization of cognitive radio networks
powered by energy harvesting,\rq\rq \emph{ Proc. IEEE INFOCOM'15}, HongKong, Apr. 27 - 30, 2015.

\end{thebibliography}
%

\end{document}